\documentclass[showpacs,preprint,preprintnumbers,amsmath,amssymb,a4paper,aps]{revtex4}
\usepackage{amsmath}

\usepackage{graphicx}

\begin{document}
\title{A model for motor-mediated bidirectional transport along an antipolar microtubule bundle}
\author{Congping Lin$^{1,2}$, Peter Ashwin$^1$ and Gero Steinberg$^2$}
\affiliation{Mathematics Research Institute$^1$ and School of Biosciences$^2$\\
University of Exeter\\
Exeter EX4 4QF}

\date{(\today)}

\begin{abstract}

Long-distance bidirectional transport of organelles depends on the
motor proteins kinesin and dynein. Using quantitative data obtained from a fungal model system, we previously developed ASEP-models of bidirectional motion of motors along unipolar microtubules (MTs) near the cell ends of the elongated hyphal cells (herein referred as ``unipolar section''). However, recent quantitative live cell imaging in this system has demonstrated that long-range motility of motors and their endosomal cargo mainly occurs along extended antipolar microtubule bundles within the central part of the cell (herein referred to as ``bipolar section''). Dynein and kinesin-3 motors coordinate their activity to move early endosomes (EEs) in a bidirectional fashion, with dynein mediating retrograde motility along the unipolar section near the cell poles, whereas kinesin-3 is responsible for bidirectional motions along the antipolar section. Here we extend our modelling approach to simulate bidirectional
motility along an antipolar microtubule bundle. In our model, cargos (particles) change direction on each MT
with a turning rate $\Omega$ and the MTs are linked to each other at
the minus ends where particles can hop between MTs with a rate $q_1$
(obstacle-induced switching rate) or $q_2$ (end-induced switching
rate). By numerical simulations and mean-field approximations, we
investigate the distribution of particles along the MTs for
different overall densities $\Theta$. We find that even if $\Theta$
is low, the system can exhibit shocks in the density profiles near
plus and minus ends caused by queueing of particles. We also discuss
how the switching rates $q_{1,2}$ influence the type of motor that
dominates the active transport in the bundle.
\end{abstract}

\pacs{87.10.Mn, 87.10.Hk, 87.16.Wd}

\maketitle

\section{Introduction}

Spatial redistribution of organelles is of central importance to all
eukaryotic cells. Long-distance transport involves the activity of
molecular motors that move along polymers of tubulin dimers, the
so-called microtubules (MTs), powered by the hydrolysis of ATP
\cite{Vale2003}. Bidirectional transport along MTs involves the
opposing motor molecules kinesin and dynein where kinesin takes
cargos to the plus ends of MTs and dynein takes the cargo to the
minus-ends \cite{Gross2004}.

Numerous theoretical studies attempt to describe motility behaviour
of cargos along a single MT or a network of MTs, e.g., see
\cite{Nishinari2005,Hough2009,Campas2008}. One of the simplest and
best studied models is the {\em asymmetric simple exclusion process}
(ASEP) defined on a single track; see review~\cite{Blythe2007} and references therein. Extended ASEP
models for bidirectional transport have made various assumptions to
avoid collisions between opposite-directed particles. One is
assuming that particles are binding and unbinding to/from tracks
\cite{Lipowsky2001,Klumpp2003,Ebbinghaus2009,Ebbinghaus2010}.
Another is assuming that the exclusion principle only applies to
particles moving in the same direction and presence of a motor in
the opposite direction modifies the rate at which motors enter into the
site \cite{Liu2010}. Alternatively, a high direction-change
rate can avoid clusters due to collision \cite{Muhuri2011}. Moreover, Evans {\em et al} introduced another possibility~\cite{Evans1995} to avoid collisions by allowing particle interchanges when they meet. Juh\'asz introduced a two-lane ASEP \cite{Juhasz2007} with
opposite-directed particles moving in separate lanes, thereby avoiding collisions between opposite-directed particles. The two-lane ASEP is equivalent to a two-species ASEP in some sense as discussed in \cite{Reichenbach2007}. We introduced a multilane model in \cite{Lin2011} where particles can
change protofilaments to avoid collisions, taking into account that a
single MT consists of 13 protofilaments \cite{Tilney1973}, each of which provides a
potential track for motors.

Based on recent advances in live cell imaging techniques and the use
of the fungal model system {\em Ustilago maydis}, {\em in vivo}
observation of dynein indicates that collision between
opposite-directed motors rarely occurs
\cite{Schuster2011a,Schuster2011b}. This allows adapting the
two-lane model in \cite{Juhasz2007} to investigate the bidirectional
transport of dynein motors on unipolar MTs. The adapted two lane
model together with a more sophisticated 13-lane model provides an
explanation for the formation of dynein accumulation at MT
plus-ends \cite{Ashwin2010, Schuster2011a,Lin2011}. This dynein
accumulation at MT plus ends is suggested to prevent the cargo -
early endosomes (EEs) - falling off the MT \cite{Schuster2011a}.
More recent work on {\em U. maydis} has shown that the majority of
the fungal cell contains antipolar MT bundles and that unipolar MTs are restricted to the cell poles \cite{Schuster2011c}. Early endosomes - the main cargo of dynein and kinesin-3 in hyphal cells
\cite{Lenz2006,Wedlich-Soldner2002}, undergo long-distance
bidirectional motility \cite{Wedlich-Soldner2000}. Interestingly,
bidirectional long-distance motility of EEs along the bipolar MTs is
mainly mediated (dominated) by kinesin-3, whereas dynein is
mediating retrograde motility of EEs along the unipolar MT~\cite{Schuster2011c}. That paper also shows that EEs travel over the entire length of the MT array and concludes that dynein and
kinesin-3 cooperate and that cargo can hop between MTs within the
bundle \cite{Schuster2011c}. They also show that during the EE
transport, dynein can detach from the cargo and the EE continues
motility (carried by kinesin-3) after a short pause. Moreover, short
pauses of EEs before they continue the directed motility are
observed at minus ends of MTs by visualizing both minus ends of MTs and EEs
\cite{Schuster2011c}. These suggest that EEs may change MTs at MT
minus ends by altering their active motor types.

Hopping between tracks has been modelled for
unidirectional traffic
\cite{Brankov2004,Pronina2005,Wang2008,Embley2009,Neri2011} where a single track is followed by parallel tracks. The junction between single and parallel tracks allows particles on the single track to step into either of
the parallel tracks. Here we provide a new model that is based on our previous ASEP model but which includes this junction mechanism to describe bidirectional motility of cargo along an antipolar
microtubule bundle. In our model, the lattice is composed of two MTs
that are coupled at microtubule minus ends and the arrangement of
antipolar MT bundle gives unipolar and bipolar sections within the
bundle. A detailed description of the model is given in
Section~\ref{sec_bipolarmodel}. In Section~\ref{sec_meanfield}, we
show that the distribution of particles along the lattice can
exhibit a variety of phases depending on the parameters; in
Subsection~\ref{sec_unipolar} the phases in the unipolar section are
analyzed while in the following subsections the phases in the
entire system are considered. In particular, we show that even at a
low overall density of particles, particles can accumulate at minus
ends as well as at plus ends. Moreover, we find a novel type of phase where density profiles of one type of particles can smoothly connect between low and high density. The density profiles for these different phases are
well approximated by our mean field analysis. In
Section~\ref{sec_domin}, we consider how the MT switching rates $q_{1,2}$ affect
the contribution of each type of motors to the transport of cargos.
Finally, in Section~\ref{sec_dis} we discuss the biological relevance
of this model.

\section{A lattice model with antipolar bundling}\label{sec_bipolarmodel}

\begin{figure}
  \centering
  \includegraphics[width=14cm]{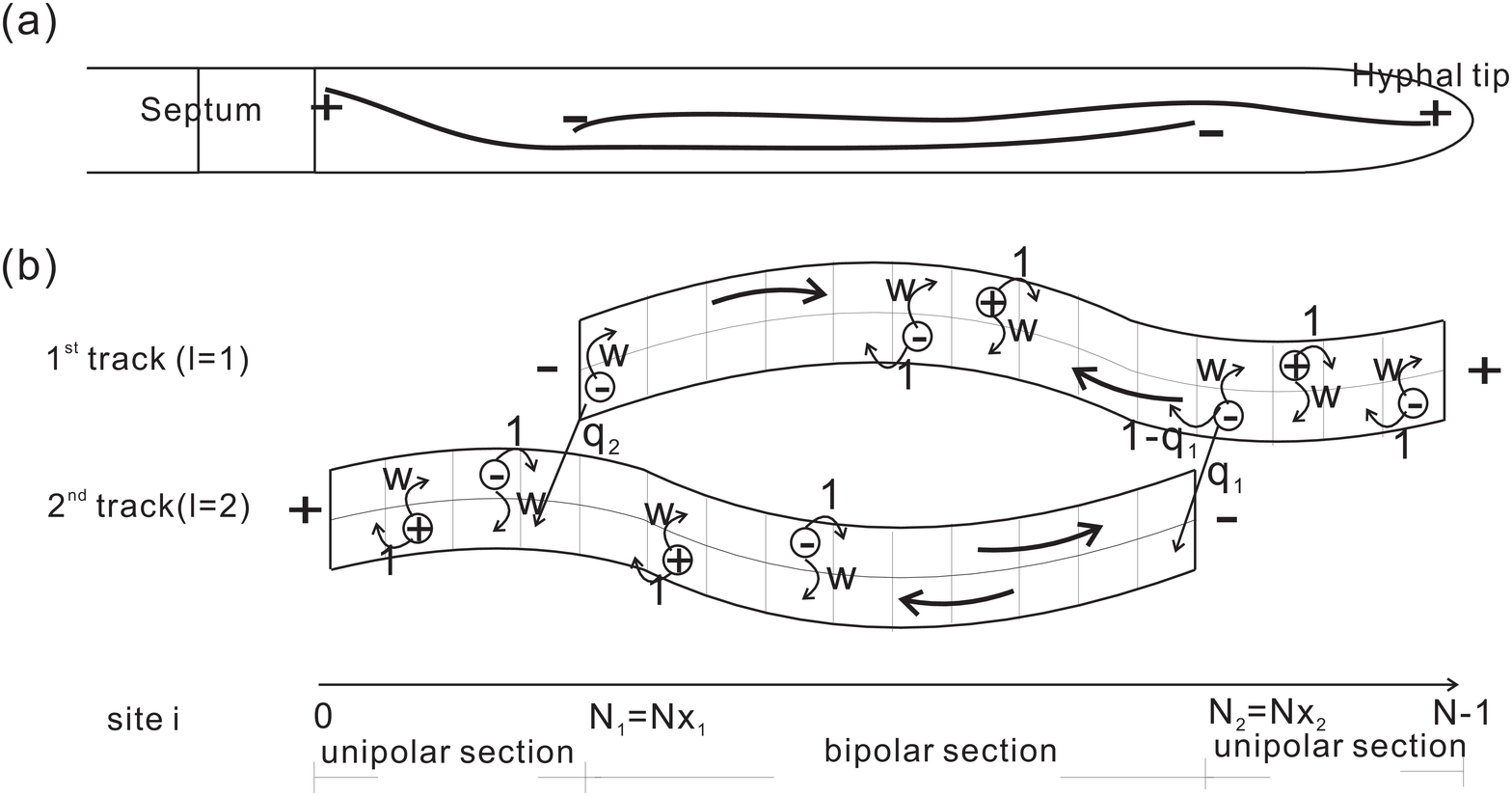}\\
  \caption[Schematic diagram of ASEP on an antipolar MT bundle]{(a) shows a schematic hypha where two MTs with plus ends at cell poles (tip and septum) form an antipolar bundle. (b) shows a schematic diagram of bidirectional transport on the bundle which is presented as a discrete lattice of length $N$. Two tracks in the lattice represent two MTs and are of length $N-N_1$ and $N_2$ respectively with plus ends at two ends of the lattice. Plus- and minus-type particles on each track move on separate lanes with indicated orientation. The forward rate on the same track is set to be 1 except at the junctions between the {\em bipolar section} and the {\em unipolar sections} for minus-type particles. Particles can also change directions on the same track with rate $\omega$. We assume that track switching only occurs at the junctions between sections for minus-type particles and occurs with rate $q_{1,2}$. If $q_{1,2}=0$ then the two MTs are uncoupled.}\label{fig_MT}
\end{figure}

In this section, we introduce a simple discrete-lattice model with antipolar bundling
of two MTs (we refer to each MT as a track). The bundle is presented by a lattice of length $N$ and contains two
tracks of length $N-N_1$ and $N_2$ respectively that overlap in a common section of length $N_2-N_1$. The two plus ends of MTs are located at two ends of the
lattice and the other ends of two MTs - the minus ends (marking the end of the overlap) - are in the middle; see Figure~\ref{fig_MT}. The middle section is referred to as the {\em bipolar section}. The left sections in the bundle are referred to as the {\em unipolar sections} which are of a relative length $x_1$ and $1-x_2$ respectively where $x_{1,2}:=N_{1,2}/N$. Considering the symmetric organization of the MT bundle shown in \cite{Schuster2011c}, we assume a symmetric lattice with $x_1+x_2=1$ in the following.

In the bundle, each track can support bidirectional transport of
particles (particularly EEs here) driven by opposite-directed motors
(kinesin-3 and dynein); these particles are of two types: plus- and
minus-type. Plus-type particles are driven by plus-directed motors
moving towards the plus end of the track whereas vice versa for
minus-type particles. Reversal of transport direction along each
track can be realized by a brief ``tug-of-war'' event
\cite{Mueller2008,Mueller2010,Hendricks2010,Soppina2009} between
counteracting motors on the particle. As rare collisions between
opposite directed motion of EEs in {\em U. maydis} are observed
\cite{Schuster2011c} which is similar to the dynein transport, we
assume transport on each track in a particular direction is on a
separate lane as in the two-lane model \cite{Juhasz2007,Ashwin2010}
for convenience. Thus, the lattice contains four lanes, each of
which supports one direction and a single type of particles as
illustrated in Figure~\ref{fig_MT}~(b). As EEs rarely fall off the
MTs \cite{Schuster2011a,Schuster2011c} in the {\em U. maydis}, we
assume the transport system is closed, i.e., there is no injection or exit of
particles into/out of the lattice at two ends of each track. The closeness of the system
leads a particle number conservation during the transport. Particles once reach the end of tracks will wait until they change direction to continue the motility.

Further assumptions on the model are made on track switching of
particles based on experiment observation of EE motility around
minus ends in \cite{Schuster2011c}. An antipolar bundle of tracks
enables particles to switch between tracks together with a change of type and without reversing direction. This form of track switching has been observed in {\em U. maydis} \cite{Schuster2011c}. Track switching
may be possible inside the bipolar section as well as at the
junction between bipolar and unipolar sections. For simplification, in this paper we
assume track switching occurs when passing the junctions associated with minus ends of MTs. Due to the
bipolar organization in the bundle, plus-type particles can only be
on the same track when crossing the junctions. When a minus-type
particle moves from a unipolar to a bipolar section, the minus end
in the junction may enhance plus-directed motors on the particle to
hop onto the other track, resulting in a change of the track on
which the particle moves and the type of the particle. Thus the
minus ends can be viewed as ``obstacles'' for minus-type particles
when attempting to step forward on the same track. This assumption on rack switching is sufficient to allow particles to
travel across the entire lattice from one plus end to the other.


In our model, we identify each location in the lattice by a pair
$(l,i)$ together with a transport direction (corresponding to a lane in
each track $l$) where $l\in\{1,2\}$ and $i\in\{0,\cdots,N-1\}$
denotes the site along the lattice. Particles hop from one location
to another with a certain transition rate where the possible transitions we
consider are listed below.

\begin{description}

\item {\bf Forward.} Plus and minus-type particles on each location can move forward to the nearest site along the same lane. We assume equal forward rates for plus- and minus-type particles (corresponding to same velocities which is suggested from {\em in vivo} transport~\cite{Schuster2011a,Schuster2011b}) and homogeneous forward rates on each lane except at the junctions $(1,N_2)$ and $(2,N_1-1)$ for minus-type particles. For convenience, we let the equal forward rate $p=1$ and other rates in units of $p$. For the forward rate between junctions for minus-type particles, we set the rate to be $1-q_1$ where $q_1$ is the
obstacle-induced switching rate given below.

\item {\bf Track switching.} We allow minus-type particles at junctions $(1,N_2)$ and $(1,N_1)$ to step forward onto the second track together with a type change with rate $q_1\in[0,1]$ (called obstacle-induced switching rate) and $q_2\in[0,1]$ (called end-induced switching rate) respectively. Similarly, minus-type particles on the second track can switch onto the first track at the junctions with the same obstacle-induced/end-induced rate.

\item {\bf Direction change.} Plus- and minus- type particles can change directions on the same track and the same site (by changes in type/lane) along the entire lattice. We assume homogeneous and equal direction-change rates for plus- and minus-type particles which are denoted by $w$. In addition, we define $\Omega:=\omega N$ which characterizes the overall direction-change rate.
\end{description}

The forward and obstacle-induced switching rates for minus-type
particles at junctions $(1,N_2)$ and $(2,N_1-1)$ are chosen to
preserve the overall forward rate $p=1$. All transitions described
above are subject to a simple exclusion principle, i.e., there can
be at most one particle at each location of the lattice, and
presence of a particle at one location prevents other particles from
moving into that location.

In the time evolution of this stochastic bidirectional transport,
the occupancy at each location of the lattice for plus- and
minus-type particles at time $t$, $\tau^{l}_{\pm,i}(t)$ changes
according to the above transition rates (assumed to take place
independently and instantaneously) from an initial state
$\tau^l_{\pm,i}(0)$. The exclusion principle ensures that
$\tau^{l}_{\pm,i}(t)\in\{0,1\}$. A special case of the model is
$q_1=q_2=0$ where the transport on each of the two tracks is
independent of the other. In this case the process is no longer
ergodic. For the ergodic cases, the statistically stationary state will be
independent of initial conditions and a symmetric lattice structure
gives a symmetric distribution of particles on two tracks where the
total numbers of particles are equal. In order to compare transport
properties with different switching rates $q_{1,2}$, we assume an initial condition
satisfying that each track possesses an equal number of particles
and assume that the system has reached a statistically steady state.

In statistically stationary state, densities of plus- and
minus-type particles on each location $(l,i)$ are defined as the mean
occupancy of the particles:
$$\rho^l_i:=\langle\tau^l_{i,+}\rangle,~\sigma^l_i:=\langle\tau^l_{i,-}\rangle$$ where the brackets $\langle\cdot\rangle$ denotes the ensemble average. These densities are related to the overall density $\Theta$ of particles expressed as
\footnote{In the unipolar section, say $i<N_1$, only the location
$(2,i)$ is included in the defined lattice; for convenience we set
$\tau^1_{\pm,i}=0,~i<N_1$ and so for another unipolar section.}
$$\Theta=\Theta_++\Theta_-=\frac{\sum_{i=0}^{N-1}(\rho^1_i+\rho^2_i)}{2(N-N_1+N_2)}+\frac{\sum_{i=0}^{N-1}(\sigma^1_i+\sigma^2_i)}{2(N-N_1+N_2)}$$
where $\Theta_{\pm}$ are the overall densities of plus- and minus-type
particles respectively. The overall density $\Theta$ is conserved under the time evolution whereas $\Theta_{\pm}$ may not be conserved. Meanwhile, the mean currents (the mean rate of stepping forward within unit time) away from the junctions are given by
$$
J^1_{\pm,i}=\langle\tau^1_{\pm,i}(1-\tau^1_{\pm,i\pm 1})\rangle
$$ in units of
$p$ on the first track. At the junctions between sections, the mean currents for minus-type particles
are
$$
J^1_{-,N_2}=(1-q_1)\langle\tau^1_{-,N_2}(1-\tau^1_{-,N_2-1})\rangle+q_1\langle\tau^1_{-,N_2}(1-\tau^2_{+,N_2-1})\rangle
$$
and
$$
J^1_{-,N_1}=q_2\langle\tau^1_{-,N_1}(1-\tau^2_{+,N_1-1})\rangle.
$$
Similarly, we can write the expressions for the mean currents on the second track. The net
current in the unipolar section is zero and at the junction between
sections it is balanced: $J^1_{-,N_2}=J^1_{+,N_2-1}+J^2_{-,N_2-1}$.

The exact analytical solutions of the density profiles in our model
with general parameters can be very difficult or impossible to find using methods such as matrix production \cite{ Blythe2007} or Bethe ansatz \cite{Derrida1999} and
therefore we use mean-field approximations and numerical
simulations. For the continuous-time discrete-state model, we use a
Gillespie algorithm \cite{Gillespie1977} to simulate the time
evolution. The parameters that govern this bidirectional transport in {\em U. maydis} are not yet fully known from experiments, but we do set some parameters informed by known properties of early endosome transport in {\em U. maydis}. We use
$x_1=1-x_2=0.2$, as microtubule minus ends in {\em U. maydis} hyphae
are shown to be approximately uniformly distributed in the middle at
about 10\% in hyphal length away from cell poles where the MT minus
ends are ``almost'' absent \cite{Schuster2011c}. For the parameter
$\Omega$, {\em in vivo} experiments in
\cite{Schuster2011a,Schuster2011b} suggests a range of run length
$M=10\sim70~\mu m$ in a hyphal length of $L=100~\mu m$ which
gives \footnote{The unit $p$ is given by $p=v/h$ where $v$ is the
velocity and $h$ is the space step and the turning rate
$\omega=v/M$. Hence the quantity $\Omega$ in unit of $p$ is given by
$\Omega=\omega N/p=L/M$.} $\Omega\in[1,10]$. A lattice
length of $N=500$ is used for simulations unless otherwise
stated.

\subsection{Symmetry of the system}\label{sec_symm}

The standard unidirectional ASEP on a single lane and the two-lane
ASEP developed by Juh\'asz \cite{Juhasz2007} share a common feature
- particle-hole symmetry. This feature does not hold in the model we
introduce here. However, by adapting the
forward stepping rate of plus-type particles when crossing the
junctions between sections (i.e., assuming there is inhomogeneity in
stepping rates for plus-type particles at locations $(1,N_2-1)$ and
$(2,N_1)$), the model can still exhibit an exact particle-hole
symmetry as explained below. Note that particles are of two types,
either moving towards or away from the plus ends on each track, we
divide holes in the lattice into two types. Minus-type holes
refer to the holes in lanes for plus-type particles (i.e., the first and last lanes in the lattice shown in Figure~\ref{fig_MT}), as these holes move towards the minus end when the corresponding plus-type particles
step forward; plus-type holes refer to the holes in lanes for
minus-type particles. An (obstacle-induced) track switching of
minus-type particle from the unipolar to bipolar section indicates
an (end-induced) track switching of a plus-type hole from the bipolar
to unipolar section. Similarly, a plus-type particle stepping from the
bipolar to unipolar section on the same track indicates that a minus-type hole steps
from the unipolar to bipolar section. Therefore, if we let the forward rate for plus-type particles when crossing the junction at locations $(1,N_2-1), (2,N_1)$ be $\hat{p}$ and choose $\hat{p}=1-q_2$, then the
system possess the particle-hole symmetry, i.e., the system is identical
under the exchange:
$$
\hat{p}\leftrightarrow 1-q_1,~~q_1\leftrightarrow q_2,~~\tau^l_{\pm,i}\leftrightarrow
1-\tau^l_{\mp,i}.
$$
The uncoupled system with homogeneous rates for plus-type particles (i.e., $q_{1,2}=0$ and $\hat{p}=1$) is a special case with particle-hole symmetry. For small $q_1$, the system with homogeneous rates for plus-type particles (i.e., $\hat{p}=1$) still have some approximately ``symmetric'' behaviour. Thus we focus on low and intermediate overall densities; information for high overall densities can then be partly deduced from this ``symmetry''.

\section{Steady state distribution of particles and mean-field analysis}\label{sec_meanfield}

In the lattice of length $N$, we take a rescaled position variable
defined as $x = i/N\in[0,1]$. In the continuum limit where $N\to
\infty$, we re-express $J^l_{\pm}(x)$ as the mean unidirectional current to the plus/minus end on each track ($l=1,2$) and $J^l(x):=J^l_+(x)-J^l_-(x)$ as the net (mean) current. Moreover, we re-express the densities on each track as $\rho^l(x)$ and
$\sigma^l(x)$ for plus- and
minus-type particles respectively. The overall density of plus- and
minus-type particles can then be re-expressed as
$$
\Theta_+=\frac{\int_0^1(\rho^1(x)+\rho^2(x))dx}{2(1-x_1+x_2)},~~\Theta_-=\frac{\int_0^1(\sigma^1(x)+\sigma^2(x))dx}{2(1-x_1+x_2)}.
$$
In the spatial symmetric lattice with $x_1+x_2=1$, we have
$\rho^1(x)=\rho^2(1-x)$ and $\sigma^1(x)=\sigma^2(1-x)$ for these
well defined regions. Hence we only need to consider the
distribution of particles on one track, say the first track, and can
ignore the track index superscripts without ambiguity.

\begin{figure}
  \centering
  \includegraphics[width=14cm]{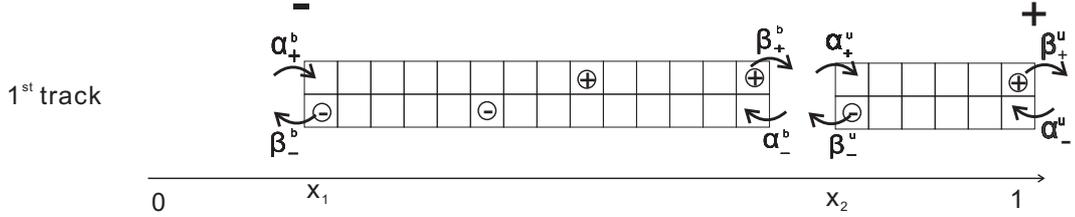}\\
  \caption{One of the tracks for ASEP with antipolar bundling described in Figure~\ref{fig_MT} is mapped onto two unbundled bidirectional ASEPs coupled by boundary conditions as shown here. This shows the sections of the first track. The boundary rates $\alpha_-^u=\beta_+^u=0$ as the system in Figure~\ref{fig_MT} is closed.}\label{fig_eff_BC}
\end{figure}

As the lattice is composed of unipolar and bipolar sections, we
consider these sections separately with appropriate boundary rates $\alpha^{u(b)}_{\pm},\beta^{u(b)}_{\pm}$;
see Figure~\ref{fig_eff_BC}. The stochastic process in each section
is an ASEP similar to that discussed in \cite{Juhasz2007,Ashwin2010} but with different boundary conditions. In both ASEPs in \cite{Juhasz2007,Ashwin2010},
the mean-field approximation predicts the density profiles well for statistically stationary states. For the mean-field approximation, we ignore
two-point correlations. The density profile in the steady state
for plus- and minus-type particles in the unipolar/bipolar section
of the lattice is then governed by the following equations (ignoring the
second  derivative and second order of $1/N$) with appropriate
boundary conditions (see \cite{Juhasz2007,Ashwin2010} for details)
\begin{eqnarray}\label{eq_pdesimple}
0 &=& (2\rho-1) \frac{d\rho}{dx} - \Omega(\rho-\sigma)\\
0 &=& (1-2\sigma) \frac{d\sigma}{dx} +\Omega(\rho-\sigma).\nonumber
\end{eqnarray}
Meanwhile, the unidirectional currents within each section are given by equations
$$J_+(x)=\rho(x)(1-\rho(x)),~~~J_-(x)=\sigma(x)(1-\sigma(x))$$
and the net current reads as
$$J(x)=\rho(x)(1-\rho(x))-\sigma(x)(1-\sigma(x)).$$

Taking the sum of equations in (\ref{eq_pdesimple}) gives
\begin{equation}\label{eq_netcurrent}
\frac{dJ(x)}{dx}=
\frac{d}{dx}\left(\rho(1-\rho)-\sigma(1-\sigma)\right)=0
\end{equation}
which gives a constant net current in the mean field approximation
$J(x)=J_0$ within each section in the steady
state. A positive (negative) net current indicates a net current towards the plus (minus) end.

\subsection{Spatial distribution in the unipolar section}\label{sec_unipolar}
The system we consider is closed, and so the unipolar sections are half closed, i.e., no exit (of plus-type particles) and no injection (of minus-type particles) is possible at the plus ends. For the first
track, the unipolar section in $x_2<x\leq 1$ is closed at the right
end, corresponding to the boundary rates $\alpha^u_-=\beta^u_+=0$, which give
$$
\rho(1)=1,~\sigma(1)=0.
$$
We assume the left end $x=x_2$ of the unipolar section is associated with injection/exit rates
$\alpha^u_+$ and $\beta^u_-$. As discussed in~\cite{Ashwin2010}, the
unipolar section has a zero net current due to the closed boundary
at the right end. Thus
$J=\rho(x)(1-\rho(x))-\sigma(x)(1-\sigma(x))=0$, leading to the
density relation
$$\rho(x)=1-\sigma(x),~~\rho(x)=\sigma(x).$$
Together with the ODE~(\ref{eq_pdesimple}), we have two possible solutions - complementary density
\begin{equation}\label{eq_comp}
\rho(x)=1-\sigma(x)=1-\Omega+\Omega x
\end{equation}
and equal density $\rho(x)=\sigma(x)=C$. The equal density may
appear in the bulk of the unipolar while the complementary
density starts from the plus end and extends toward the interior of
the unipolar section.

\begin{figure}
  \centering
  \includegraphics[width=14cm]{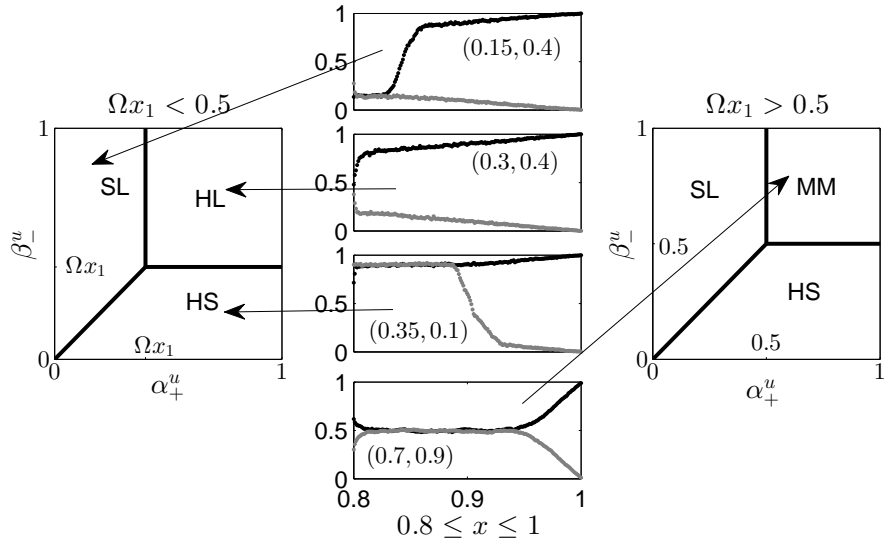}\\
  \caption[Phase diagram of the density profile in the unipolar section]{Phase diagram of the density profile in the unipolar section with general left boundary conditions $\alpha^u_+$ and $\beta^u_{-}$ and closed right boundary condition for both cases $\Omega x_1<1/2$ (left panel) and $\Omega x_1>1/2$ (right panel), where $x_1=1-x_2=0.2$. Numerical examples of density profiles in the unipolar section for each phase are illustrated with indicated boundary conditions $(\alpha^u_+,\beta^u_+)$. The top three density profiles in the middle column are using $\Omega=1$ while the bottom density profile is using $\Omega=10$. In the density profiles, black dots are for plus-type particles while gray dots are for the minus-type.}\label{fig_phase_twolane}
\end{figure}

As discussed in \cite{Ashwin2010}, for low injection and high exit
rates, the constant in the equal density case reads as
$C=\rho(x_2)=\alpha^u_{+}$; the density profile of plus-type
particles exhibits a shock between equal density and
complementary density regimes while the density of minus-type
particles is continuous except near the left end. We refer this as
an SL phase; the first letter means for the plus-type particles while the second one for the minus-type particles; S stands for shock while L stands for low (less than a half). The shock location $x_s$ is determined by matching
$\lim_{x\to x^-_s}\rho(x)=\lim_{x\to x_s^+}1-\rho(x)$. Combining the
complementary solution~(\ref{eq_comp}) and equal density
$\alpha^u_+$ gives $x_s=1-\frac{\alpha^u_+}{\Omega}$. A limit
$\lim_{x_s\to x_2}\alpha^u_+(x_s)=\Omega x_1$ gives a boundary of
this SL phase in the space $(\alpha^u_+, \beta^u_-)$. When the shock
is driven out of the unipolar section, the plus-type particles are
in high density (over one half) and minus-type particles are in low
density; we refer as an HL phase; H stands for high density. The linear density profile of
plus-type particles in the complementary density decreases from
1 at the plus end and propagates inside. When a lower bound of $1/2$
for plus-type particles in density is reached before $x$ is
decreased to $x_2$, the density of $1/2$ continues when $x$ is
further decreased and the maximum unidirectional current occurs in
the unipolar section. This is referred as an MM phase (M stands for maximal unidirectional current, which is $1/4$) and only
occurs when $1-\Omega+\Omega x_2<1/2$, i.e., $\Omega x_1>1/2$. By
particle-hole symmetry, analogous to the SL phase, we have an HS
phase where the equal density is over one half and minus-type
particles experience a shock. In summary, the phase diagrams of
density profiles for the unipolar section with general left boundary
conditions $\alpha^u_+$ and $\beta^u_-$ are shown in
Figure~\ref{fig_phase_twolane} for both $\Omega x_1<1/2$ and $\Omega
x_1>1/2$. In the coexistence line $\alpha^u_{+}=\beta^u_{-}$, we
have $\rho(x_2)=\alpha^u_+=1-\sigma(x_2)$ and note that
$\rho(1-\frac{\alpha^u_+}{\Omega})=1-\alpha^u_+$ in the
complementary density, thus a similar discussion as in
\cite{Juhasz2007} can be applied in the region $[x_2,
1-\frac{\alpha^u_+}{\Omega}]$.

\subsection{Analysis of spatial distribution along the entire bundle}\label{sec_den_entire}
In the following subsections, we aim to analyze the density profiles
in the entire symmetric bundle by considering two ASEPs on the first
track with corresponding boundary rates in each section. The
boundary rates in one section may depend on those in another
section.

In the uncoupled case ($q_1=q_2=0$), the stochastic process on
each track is exactly the same as the two-lane ASEP discussed in
\cite{Juhasz2007} assuming equal direction-change rates. Also from the
discussion of the unipolar section in Section~\ref{sec_unipolar}, we can see that the density profile on the first track consists of three segments in
general; an equal-density segment occurs in the middle and connects
with the boundaries by complementary-density segments on its left
and right sides which are continuous at $x=x_1$ and $1$ respectively.

For either $q_1$ or $q_2$ positive, the density profiles can exhibit
a variety of phases even if the parameters $\Omega$ and
$\Theta$ are fixed. In the following, we show that a mean-field
approximation with appropriate boundary conditions for both unipolar
and bipolar sections agrees well with numerical simulations for a
variety of density profiles. The density profiles in the unipolar are discussed in Section~\ref{sec_unipolar}, thus we focus on the bipolar section in order to understand the density profile along the entire lattice. In the generic case where $q_1q_2>0$,
the net current in the bipolar section on a single track is not
necessary zero. For a small net current $J$, one can show that there are ``equal'' (approximately equal) and ``complementary'' (approximately complementary) density solutions $\sigma\approx \rho$, $\sigma\approx 1-\rho$; see Appendix~\ref{App_den} for detailed explanations. Particularly, when neither $\rho$ nor $\sigma$ is close to one half, we approximate the constant net current
$J=\rho(1-\rho)-\sigma(1-\sigma)$ by
\begin{equation}\label{eq_netJ}
J=\mbox{sgn}(1-2\rho)(\rho-\sigma) \mbox{~and~}
J=\mbox{sgn}(1-2\rho)(\rho+\sigma-1)
\end{equation}for ``equal'' and
``complementary'' densities respectively. The
solutions of density profiles for plus-type particle from~(\ref{eq_sol}) can thus be approximated as
\begin{equation}\label{eq_rho_smallJ}
\rho=-J\Omega x+C_e,~~\rho=\Omega x+C_c
\end{equation}
for ``equal'' and ``complementary'' densities respectively and the
corresponding densities of minus-type particles can be approximated as
\begin{equation}\label{eq_sigma_smallJ}
\sigma=\rho+\mbox{sgn}(2\rho-1)
J,~~~\sigma=1-\rho-\mbox{sgn}(2\rho-1) J
\end{equation}

The constants $C_{e(c)}$ in~(\ref{eq_rho_smallJ}) can be seen as functions of the boundary rates $\alpha^b_{\pm}$ and $\beta^b_{\pm}$. These boundary rates are associated
with the parameters $q_{1,2}$ in the model. Note that for a minus-type particle crossing the junction $x=x_2$ from the unipolar to the bipolar section, it switches to the other track with rate $q_1$ (which contributes to the injection of plus-type of particles on the other track in the bipolar section) and keeps on the same track with rate $1-q_1$ (which contributes to the injection of minus-type particles). Therefore, by the spatial symmetry, we approximate the
injection rates by
\begin{equation}\label{eq_BC_alpha}
\alpha^b_{+}=\sigma(x_2)q_1\mbox{~and~}\alpha^b_{-}=\sigma(x_2)(1-q_1).
\end{equation}
Moreover, a minus-type particle on the second track switches to the first track with rate $q_2$ when crossing the junction at $x=x_2$ (which contributes to the exit of minus-type particles in the bipolar section); a plus-type particle steps forward with rate 1 when crossing the junction (which contributes to the exit of plus-type particles); both the minus- and plus-type particles share the same target site when they move. Thus, by spatial symmetry, we have $\beta^b_++\beta^b_-=(1-\rho(x_2))$ and $\beta^b_-=q_2\beta^b_+$. This gives
\begin{equation}\label{eq_BC_beta}
\beta^b_{+}=\frac{1-\rho(x_2)}{1+q_2}\mbox{~and~}\beta^b_{-}=\frac{(1-\rho(x_2))q_2}{1+q_2}.
\end{equation}

\subsection{Phases for low overall densities}\label{sec_low}
For a low overall density $\Theta\ll 1/2$, the unipolar sections
have a small number of particles in occupancy, thus the unipolar is
in an SL phase where densities of both types of particles are
equally constant away from the plus end; say $\bar{\sigma}$ which is
associated with a shock location $x^u_s=1-\bar{\sigma}/\Omega$ in
the unipolar section from~(\ref{eq_comp}). When both types of particles in
the bipolar section are in low densities, they are dominated by the
injection rates $\alpha^b_{\pm}$ in~(\ref{eq_BC_alpha}) and we approximate $\sigma(x_2)$ by $\bar{\sigma}$. Thus, in the bipolar section,
the density profile of each type is governed by the parameters
$\Omega, q_{1,2}$ as well as $\bar{\sigma}$. If the end-induced
switching rate $q_2=0$, then queueing appears at minus ends in a
similar manner to the queueing at the plus end in the unipolar
section. In contrast, for a sufficiently large switching rate $q_2$,
any queuing particles at the minus end are expected to move into the
unipolar section of another track. This leads to low densities for
both types of particles in the bipolar section and we refer as an
LL-SL phase for the entire system; the two letters before the dash are for the bipolar section while the other two follow the dash are for the unipolar, e.g., here LL stands for the phase in the
bipolar while SL for the unipolar section. The explicit expressions of the mean-field approximation for the density profiles can be worked out by using ``equal'' density approximations~(\ref{eq_rho_smallJ}) and
(\ref{eq_sigma_smallJ}) for the bipolar section and considering the conservation of $\Theta$, i.e., $\Theta=\frac{\int_{x_1}^1\rho(x)+\sigma(x)dx}{2(1-x_1)}$. The detailed calculation of these expressions are given in Appendix~\ref{App_MFlow} and Figure~\ref{fig_den}~(a) shows that the mean-field approximation agrees well with numerical simulations for this LL-SL phase.

In this LL-SL phase, a boundary layer may arise near the junctions. When the boundary rate $\beta^b_-$ is satisfied, a shock of minus-type particles could form near the minus
end. We refer this case as an LS-SL phase (i.e., in the bipolar section minus-type particles are in shock state and plus-type particles are in low density, while in the unipolar section shock forms for the plus type and the density of the minus type is low); see Figure~\ref{fig_den}~(b) as an example. In the
mean-field approximation, a shock for minus-type particles in the
bipolar section stabilizes at $x=x^b_{s}$ when
\begin{equation}\label{eq_shock}
\lim_{x\to x^{b^+}_{s}}\sigma(x)=1-\lim_{x\to x^{b^-}_{s}}\sigma(x).
\end{equation}
Similar to the LL-SL phase, the density value $\bar{\sigma}$ which also
gives the boundary rate $\beta^b_-$ in (\ref{eq_BC_beta}) when approximating $\rho(x_2)$ by $\bar{\sigma}$, can in principle be worked out from the
association with parameters. Hence, we have approximations using
``equal'' and ``complementary'' densities~(\ref{eq_rho_smallJ}) and
(\ref{eq_sigma_smallJ}) for right and left sides of the shock in the
bipolar section. The detailed calculations are not shown here. Figure~\ref{fig_den}~(b) shows that the mean-field
approximation agrees well with numerical simulations for this phase.

The transition between LS-SL and LL-SL phases for
low overall density will be that the shock in the bipolar section is driven to the junction
between sections i.e., $x^b_{s}\to x_1$. In other words, the
``equal'' density of minus-type particles $\sigma(x)$ in
(\ref{eq_den_LL}) satisfies $\lim_{x\to
x_1}\sigma(x)=1-\sigma(x_1)=\beta^b_{-}$, which gives
\begin{equation}\label{eq_trans}
\frac{q_2(1-\bar{\sigma})}{1+q_2}=\bar{\sigma}q_1-\frac{\bar{\sigma}(2q_1-1)}{\Omega(x_2-x_1)+1}.
\end{equation}
Recall that $\bar{\sigma}$ is related to the overall density $\Theta$
by~(\ref{eq_thetatobar}). Figure~\ref{fig_den}~(c,d) shows examples
of this prediction on the border line between LL-SL and LS-SL phases
in the space $(q_1,q_2)$ against numerical simulations where a shock
is identified by over one half density of minus-type particles at
the last but one site to the minus end. Qualitatively, this
prediction agrees well with the simulations.

\begin{figure}
  \centering
  \includegraphics[width=14cm]{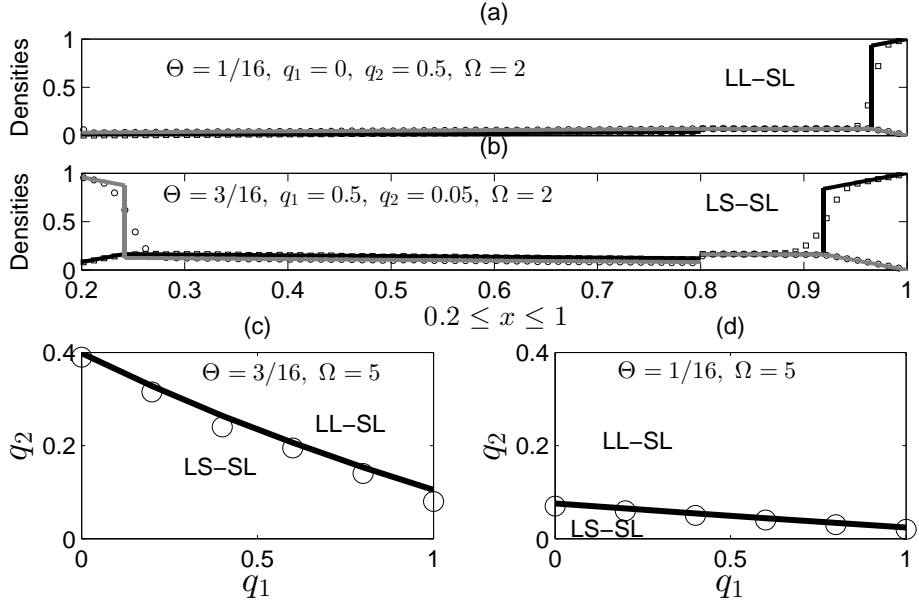}\\
  \caption{(a) and (b) show density profiles in LL-SL and LS-SL phases on the first track with $x_1=1-x_2=0.2$ and other parameters indicated. For plus-type (minus-type) particles, black (gray) lines show the density solution from mean-field approximations and squares (circles) show the averaged densities (over $T=60000~s$) from Gillespie simulations. (c) and (d) show the transition between LL-SL and LS-SL phases for indicated $\Omega$ and $\Theta$. Lines are from the mean-field approximation~(\ref{eq_trans}) while circles are from numerical simulations - a shock is identified if $\rho^1_{N_1+1}>0.5$.}
  \label{fig_den}
\end{figure}

\subsection{Phases for intermediate overall densities}\label{sec_intermediate}
For intermediate overall densities, it is clear that the system could be in the LS-SL phase and thus the HS-HS phases (via particle-hole ``symmetry''). However, there is more variety
of phases in the density profile along the entire system with different parameters.
We have not attempted to characterize all the possible phases for intermediate density, but in the following we show another two cases in details: one with shocks of both-type particles and the other with ``smooth connection'' of one type in the bipolar section.

\subsubsection{SS-HL phase}\label{sec_SS-HL}

\begin{figure}
  \centering
  \includegraphics[width=14cm]{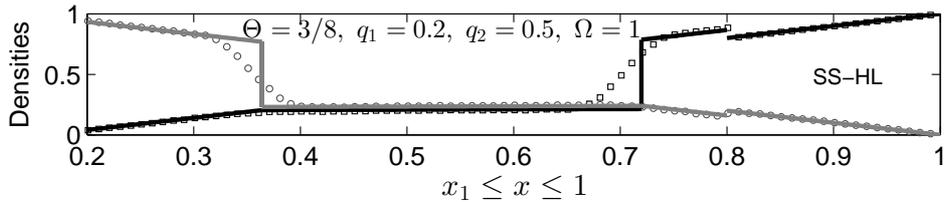}\\
   \caption{Density profile on the first track with parameters indicated shows an SS-HL phase. The markers for simulation and mean-field approximations are as in Figure~\ref{fig_den}~(a,b).}
  \label{fig_SS-HL}
\end{figure}

Note that a small $\Omega$ could lead the unipolar section to be in an
HL phase which could give shocks for both types of particles in the
bipolar section. This is referred as an SS-HL phase; see
Figure~\ref{fig_SS-HL} for an example. In this phase, the
density profiles have ``equal'' density in the middle separating two
``complementary'' density regimes in the bipolar section; the four
boundary conditions $\alpha_{\pm}^b, \beta^b_{\pm}$ given in~(\ref{eq_BC_alpha}) and~(\ref{eq_BC_beta}) are all
satisfied. Therefore, we have
$$\rho(x_1)=\alpha^b_+<1/2,~~\rho(x_2)=1-\beta^b_+,~~\sigma(x_1)=1-\beta^b_-,~~ \sigma(x_2)=\alpha^b_-<1/2$$
If $\rho$ and $\sigma$ in the bipolar section are not close to one half, then from~(\ref{eq_netJ}) we
approximate the net current by $J=\beta^b_{+}-\alpha^b_{-}=\alpha^b_{+}-\beta^b_{-}$.
We also approximate $\rho(x_2), \sigma(x_2)$ in the boundary rates by the
limit in the complementary density~(\ref{eq_comp}) as $x\to x_2$, i.e., $1-\rho(x_2)\approx\sigma(x_2)\approx x_1\Omega$. Thus, the net current is approximated
by
\begin{equation}\label{eq_J_SS}
J=\beta^b_{+}-\alpha^b_{-}=\alpha^b_{+}-\beta^b_{-}=\left(\frac{1}{1+q_2}-1+q_1\right)x_1\Omega.
\end{equation}
Similar to the discussion of the LL-SL phase, for the SS-HL phase,
the relation between the shock locations in the bipolar section and
the overall density can in principle be worked out and would give
approximated solutions of density profiles in the bundle by
using~(\ref{eq_rho_smallJ}) and (\ref{eq_sigma_smallJ}) again.
Figure~\ref{fig_deninter}~(a) shows that the approximated solution
from the mean-field agrees well with numerical simulations.

Note that the approximation~(\ref{eq_J_SS}) of the net current
suggests that the direction of the net current is governed by
$q_{1,2}$ and $(1-q_1)(1+q_2)=0$ gives a zero net current. Thus we
are expected to have constant density in a region of the bipolar
section if zero net current is satisfied as seen in Section~\ref{sec_unipolar}. Particularly, when
the overall density $\Theta=1/2$, there would be a ``maximum
unidirectional current'' region in the bulk where
$\rho^1(x)=\sigma^1(x)=1/2$. This region can be estimated by
identifying degenerated shocks (with zero shock height) for both
types of particles, which can be approximated as
$[\frac{1}{2\Omega}+(1-q_1) x_1, 1-\frac{1}{2\Omega}-q_1x_1]$. Note
also that this ``maximum unidirectional current'' could also appear with
an MM phase in the unipolar section.

\subsubsection{Phases with ``smooth connection''}\label{sec_smoothconnection}
\begin{figure}
  \centering
  \includegraphics[width=14cm]{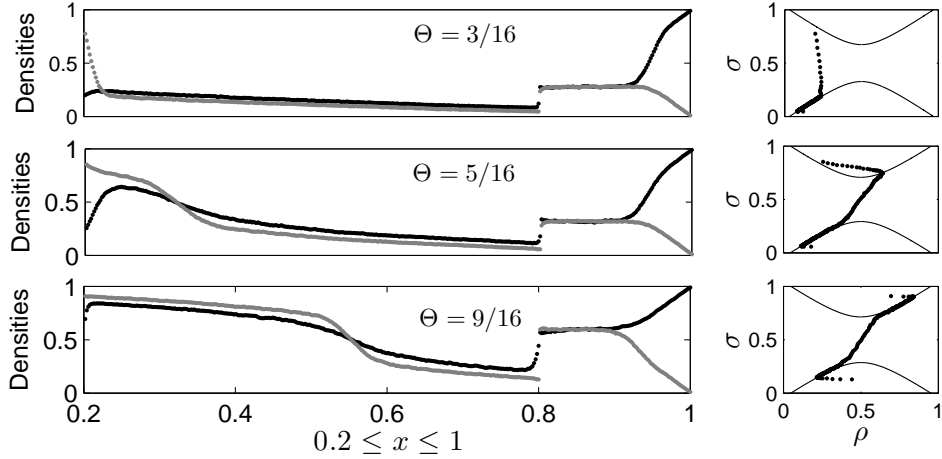}\\
   \caption{Left: density profiles on the first track for four different overall densities as indicated; dark for plus-type particles while gray for minus-type particles. Right: plot of $\rho$ vs $\sigma$ from the corresponding left panel; the solid lines show the relation $J=\rho(1-\rho)-\sigma(1-\sigma)$ where $J$ is the constant net current. Deviation of the dots from the solid lines are probably due to finite size or boundary layer effects. Other parameters are $\Omega=5, q_1=0.8, q_2=0.2$.}
  \label{fig_thetavary}
\end{figure}

\begin{figure}
  \centering
  \includegraphics[width=14cm]{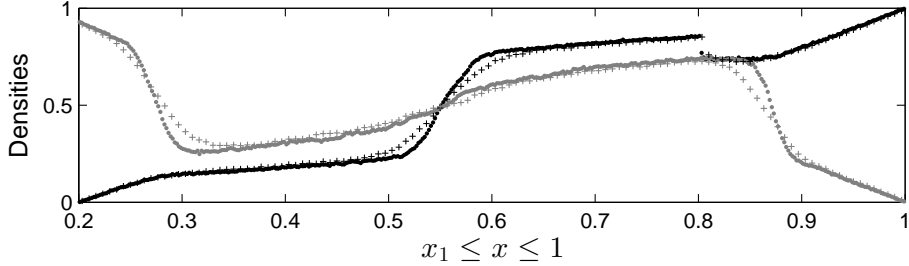}
  \caption{Density profiles on the first track for system sizes $N=300$ (`+') and $N=1000$ (`.'). The black is for $\rho$ while gray is for $\sigma$ in both system sizes. In the middle of the lattice, the shock width for $\rho$ is shortened in the larger system size; whereas $\sigma$ is almost unchanged in two system sizes. Other parameters are $\Omega:=\omega\times N=5, q_1=0, q_2=0.5, \Theta=1/2$.}
  \label{fig_size_eff}
\end{figure}

For a non-zero net current $J$, the ``equal'' density solution is a monotonic function in position seen from (\ref{eq_genODE}). Thus, simply by increasing the overall density from a low value, the maximum value of the density profiles in the ``equal'' density region, would increase and eventually reach one half. Figure~\ref{fig_thetavary} shows density profiles together with a plot of $\rho$ vs $\sigma$ under different $\Theta$; particularly, for $\Theta=5/16$ and $\Theta=9/16$, the ``equal'' density regions contain both high and low densities and one of the types (the plus-type in this example) exhibits slower change between low and high density than the other type. Moreover, Figure~\ref{fig_size_eff} compares the density profiles in different system size where the density profile in the middle is more shock-like for plus-type particles and remains almost unchanged for the other type. These suggest that in the bipolar section $\rho$ is smoothly increasing with $x$ in the ``equal'' density region while $\sigma$ has a shock profile separating low and high densities in the limiting system. We call this phase with smooth connection between low and high density for one type of particles as an SC (or a CS) phase (ignoring the existence of shocks for the type which has smooth connection in the bipolar section) depending on which type of particles exhibits ``smooth connection''; the letter `S' stands for shock while the letter `C' stands for connection. Whether the density of plus-type or minus-type particles smoothly connects low and high density is related to the sign of the net current $J$; a positive $J$ is associated with plus-type particles having smooth connection and vice versa.

\begin{figure}
  \centering
  \includegraphics[width=12cm]{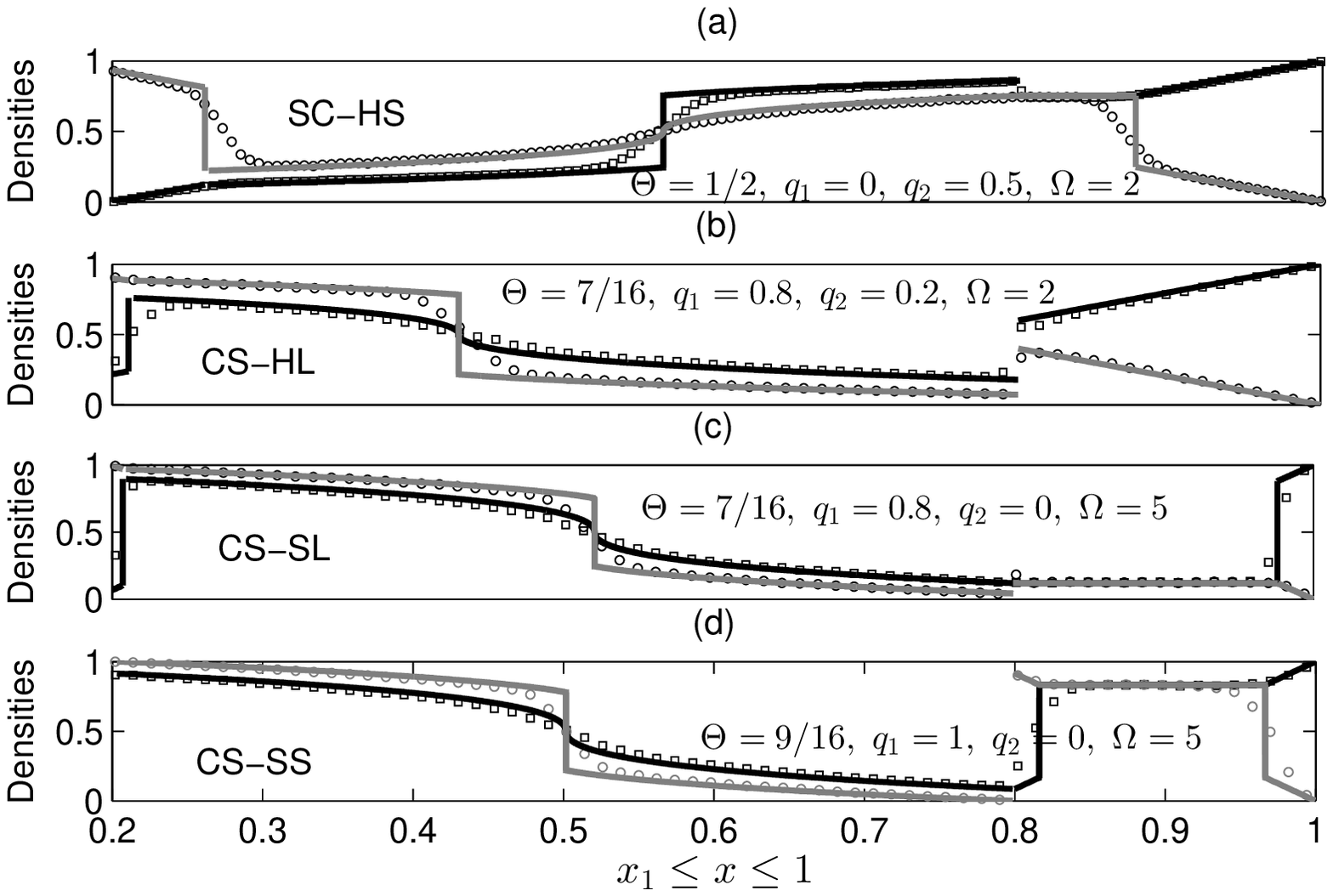}\\
   \caption{Density profiles on the first track with indicated parameters showing ``smooth connection'' phases for intermediate overall densities. A CS phase in the bipolar section indicates that minus-type particles exhibit a smooth connection through density $1/2$ at a point where plus-type particles exhibit a shock, while for an SC phase the plus-type particles exhibit the smooth connection. The markers for simulation and mean-field approximations are as in Figure~\ref{fig_den}~(a,b).}
  \label{fig_deninter}
\end{figure}

\begin{figure}
  \centering
  \includegraphics[width=12cm]{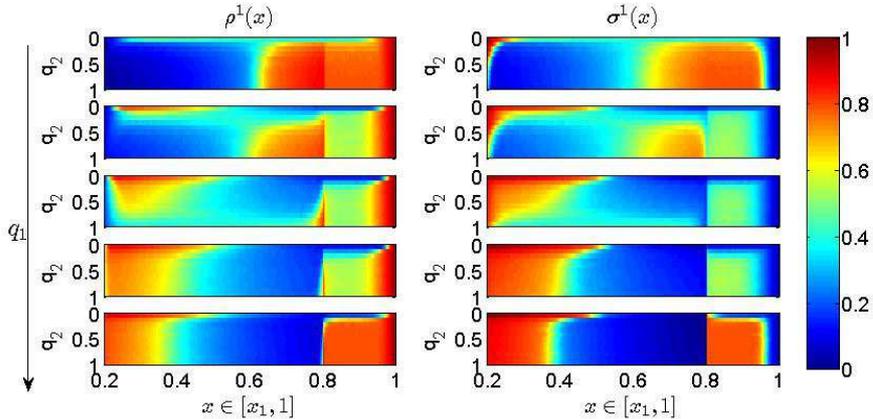}
  \caption{A variety of phases arise in the bundle for fixed $\Theta=7/16$ and $\Omega=5$ when varying the switching rates $q_{1,2}$. Each frame in the left and right panel represents densities of plus- and minus-type particles for fixed $q_1$ when varying $q_2$ by every 0.1 between 0 and 1. The fixed parameter $q_1\in\{0,0.2,0.5,0.8,1\}$ is increased from top to bottom frame in both panels.} \label{fig_den_kd}
\end{figure}

When in a CS (or an SC) phase in the bipolar section, the phase in the unipolar section can be different; see Figure~\ref{fig_thetavary} and Figure~\ref{fig_deninter} where the unipolar section can be in an HS, HL, SL or SS phase depending on the parameters. Note that the SS phase in the unipolar occurs when the boundary rate $\alpha^u_+=\beta^u_-$; in contrast to the polynomial function of density profiles shown in \cite{Juhasz2007} with symmetric open boundary condition, both density profiles exhibit shocks here due to the overall density being fixed; this is consistent with the discussion in \cite{Juhasz2007}. Furthermore, the ``smooth connection'' phases are generic in the parameter space $q_{1,2}$; seen from Figure~\ref{fig_den_kd} where each frame
represents the density profiles for $\rho(x)$ (left panel) and
$\sigma(x)$ (right panel) via color for fixed $q_1$ and changing
$q_2$ by every 0.1 between 0 and 1. The parameter
$q_1\in\{0,0.2,0.5,0.8,1\}$ is increased from top frame to the
bottom frame in both panels. In addition, there are both SC and CS phases in the parameter space $(q_1,q_2)$. For instance, plus-type particles exhibit a smooth
connection for high $q_1$ while minus-type particles exhibit for low
$q_1$.

The density profiles in ``smooth connection'' phases can also be understood by mean-field approximations. When using approximation~(\ref{eq_sol}) for densities in the bipolar section, the constants can in principle be worked out by considering the fact of the overall density conservation - similar to the previous discussion on the LL-SL phase. However, due to the non-linearity in the expression~(\ref{eq_sol}), we examine the mean-field approximation for the CS (or SC) phase by choosing appropriate $J$ and boundary conditions to satisfy the overall density $\Theta$ and the densities in the unipolar section. Take Figure~\ref{fig_deninter}~(a) as an example where minus-type particles smoothly connect low and high densities and also form a
shock in the bipolar section, and in the unipolar section densities
are in an HS phase with an equal high density which is constant away from the plus end, say $\bar{\sigma}$. In the CS-HS phase, given a net current $J$ and
$\bar{\sigma}$ with boundary conditions
$\alpha^b_{+}=\bar{\sigma}q_1$ and
$\alpha^b_{-}=\bar{\sigma}(1-q_1)$, we would have the ``equal'' and ``complementary'' densities by approximation (\ref{eq_sol}) in the bipolar section and would also have the equal and complementary densities from Section~\ref{sec_unipolar} in the unipolar section. The parameters $J$ and $\bar{\sigma}$ are chosen in order to match the shock condition for the minus-type particle in the bipolar section as
in~(\ref{eq_shock}) and the overall density $\Theta=\frac{\int_{x_1}^1\rho(x)+\sigma(x) dx}{2(1-x_1)}$. Particularly,
in this example, the mean-field solution with $\bar{\sigma}=0.753$ and $J=-0.065$ satisfy the shock condition and the overall density; Figure~\ref{fig_deninter}~(a) shows the agreements between the mean-field approximation with chosen parameters and the simulation. Similar comparisons for ``smooth connection'' phases with an HL, SL or SS phase in the unipolar section are shown in Figure~\ref{fig_deninter}~(b-d).

\subsection{Phases for high overall densities}
\begin{figure}
  \centering
  \includegraphics[width=14cm]{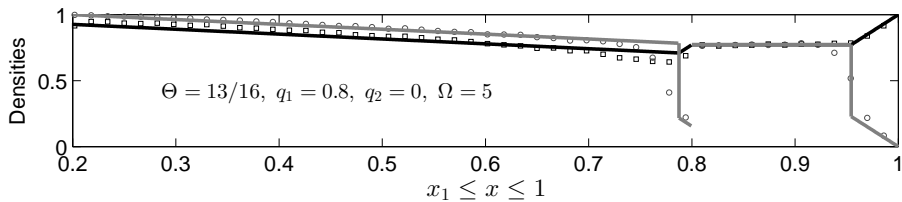}\\
   \caption{Density profiles on the first track with indicated parameters shows an HS-HS phase. The markers for simulation and mean-field approximations are as in Figure~\ref{fig_den}~(a,b).}
  \label{fig_SH-SH}
\end{figure}

When the system is in a high overall density, the vacancies are in low overall density. From the discussion of particle-hole ``symmetry'' in Section~\ref{sec_bipolarmodel}, we can see that an HH-HS and SH-HS phase could appear. Moreover, note that such ``symmetry'' may break when $q_1$ is large, in which case we find another phase where in both sections plus-type particles are in high density while minus-type particles exhibit a shock, and refere as an HS-HS phase; see Figure~\ref{fig_SH-SH}. In this phase the boundary rates $\alpha^b_-,\beta^b_{\pm}$ are satisfied and Figure~\ref{fig_SH-SH} shows that the mean-field approximation agrees with numerical simulations on the density profiles.

\section{Dominance of particles in the transport}\label{sec_domin}
Both types of particles are essential for bidirectional transport
along the bundle of tracks in the unipolar section. However, it is
possible to have only one type of particles take the bidirectional
transport across the bipolar section by track switching events and
indeed {\em in vivo} experiments of EEs suggest that kinesin-3 is
the main motor for long-range EE motility across an antipolar MT
bundle \cite{Schuster2011c}. In the following, we discuss the
contribution of each type of particles to the transport in terms of
occupancy as well as current in our model.

\subsection{Dominance in occupancy}\label{sec_occup}
For the transport in the entire bundle, one
way to quantify the dominance in occupancy is to estimate the fraction of plus-type
particles in the entire bundle, $0\leq F_+\leq 1$ where
$$
F_+:=\frac{\Theta_+}{\Theta}=\frac{\int_{x_1}^1\rho^1(x)dx}{2\Theta(1-x_1)}
$$
for a symmetric lattice $x_1+x_2=1$. Recall that $\Theta$ is an
overall density of particles in the bundle which is preserved during
the transport and is considered as a parameter. It is clear that
$F_+=1/2$ in the case $q_1=q_2=0$. Note that track switching is
accompanied with a change of type from minus to plus, thus
increasing either $q_1$ or $q_2$ will have the potential to increase
the number of plus-type particles and thus increase the fraction
$F_+$. If $F_+$ is close to $1$ then plus-type particles are in
significant dominance in terms of occupancy in the entire bundle.
For $\Omega=0$ and $q_{1,2}>0$ together with a sufficiently low
overall density $\Theta$, the maximum $F_+=1$ can be achieved.

When considering the fraction of plus-type particles within each section, it is clear that in the unipolar section there are more plus-type than minus-type particles for any parameters. In contrast, the fraction in the
bipolar section
$$
F^b_+:=\frac{\int_{x_1}^{x_2}\rho(x)dx}{\int_{x_1}^{x_2}\rho(x)+\sigma(x)dx}
$$
is less easy to estimate for general overall density $\Theta$.

\begin{figure}
  \centering
  \includegraphics[width=6cm]{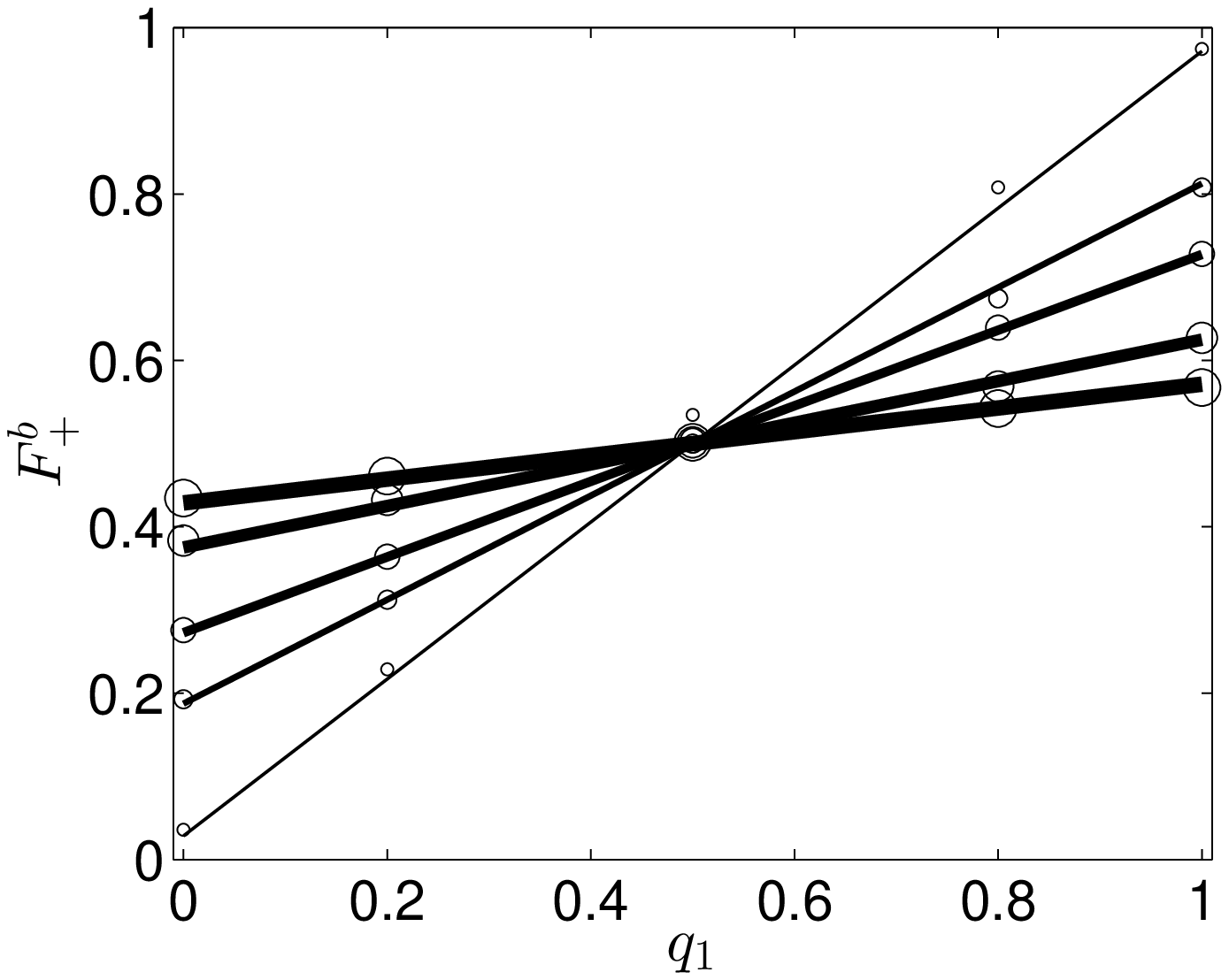}
  \includegraphics[width=6cm]{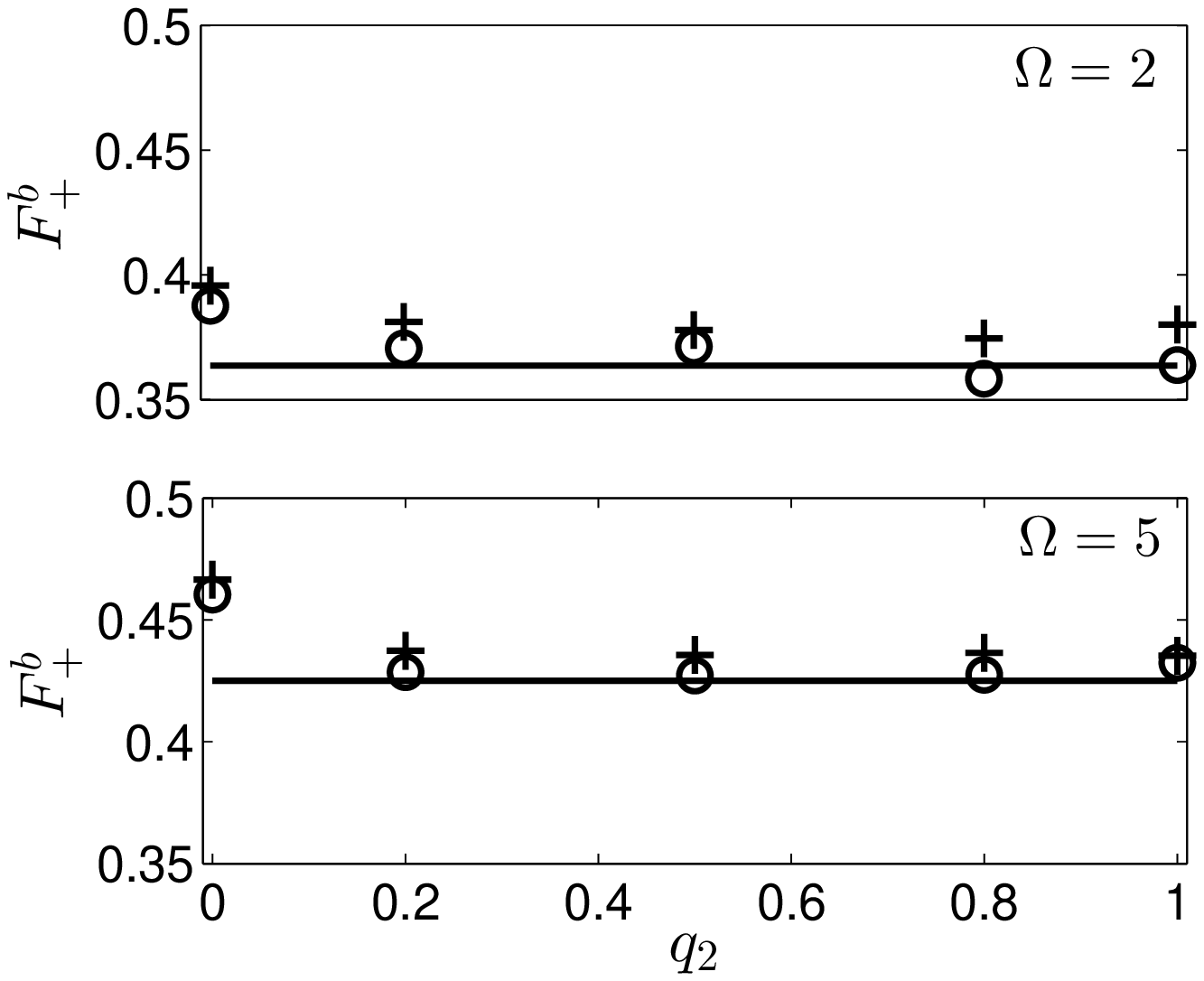}\\
  \caption{The fraction of plus-type particles in the bipolar section. Left: lines are from approximated equation~(\ref{eq_ratio_b}) for $\Omega\in\{0.1, 1, 2, 5, 10\}$ with bolder lines for larger $\Omega$ while circles are for simulations with larger circle for larger $\Omega$. Other parameters are using $\Theta=1/16$ and $q_2=1$. Right panels: to examine the independence of this fraction on the overall density $\Theta$ and switching rate $q_2$ if sufficiently large; cross and circle are for $\Theta=3/16$ and $1/16$ respectively in both up and down panels. Simulated data presented are using $q_1=0.2$ - quantitatively similar figures can be obtained for other $q_1$.}
  \label{fig_ratio_b}
\end{figure}

For low overall density where the system is in either LL-SL or
LS-SL phase, the mean-field approximation predicts the corresponding
density profiles well. Thus, for sufficiently large $q_2$ (i.e.,
larger than the critical value satisfying~(\ref{eq_trans})) which
gives low densities for both types of particles in the bipolar
section, both fractions $F_+$ and $F^b_+$ can be expressed using the
approximated density profiles, ignoring boundary layers. The
fraction in the entire bundle is
\begin{equation}\label{eq_ratio}
F_+\approx\frac{\bar{\sigma}\Omega(x_2-x_1)^2/2+\bar{\sigma}q_1(x_2-x_1)+1-\bar{\sigma}^2/(2\Omega)+(x_1-\bar{\sigma}/{\Omega})\bar{\sigma}}{\bar{\sigma}/\Omega+(1-2\bar{\sigma}/\Omega)\bar{\sigma}}
\end{equation}
where $\bar{\sigma}$ is associated with $\Theta$ as
in~(\ref{eq_thetatobar}), and the fraction in the bipolar section is
\begin{equation}\label{eq_ratio_b}
F^b_+\approx\frac{1}{2}+\frac{2q_1-1}{2\Omega(x_2-x_1)+2}.
\end{equation}
Both fractions show independence of $q_2$ and linear dependence on $q_1$. Comparatively, the latter expression for the fraction in the bipolar section is a simpler function of $\Omega$ and $q_1$ for fixed
$x_{1,2}$. We can see that this expression agrees well with simulations from Figure~\ref{fig_ratio_b}. Moreover, from this approximation, $q_1=1/2$ gives approximately equal
contribution in the bipolar section for each type of particles. This agrees with a zero net current in~(\ref{eq_J_LL}) which leads to equal density in an LL phase. Moreover, the sign of $q_1-1/2$ determines which type of particles
are in dominance and by decreasing $\Omega$, the dominance can be
enhanced. Particularly, for a small $\Omega$
$$
\lim_{q_1\to 1}F^b_+(q_1)\approx 1,~~\lim_{q_1\to 0}F^b_+(q_1)\approx 0.
$$
That is to say, for sufficiently low $\Omega$, plus-type particles are in significant dominance in the bipolar section for sufficiently large switching rate $q_1$; in contrast, minus-type particles are in significant dominance for sufficiently small $q_1$.

\begin{figure}
  \centering
  \includegraphics[width=10cm]{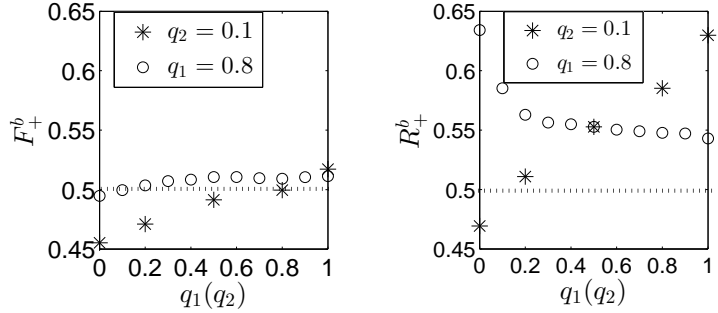} \\
  \caption[The fraction and the ratio of average current for plus-type particles]{The fraction of plus-type particles $F^b_+$ (left panel) and the ratio of average current for plus-type particles $R^b_+$ (right panel) in the bipolar section as varying $q_1$ (marked as $\ast$) or $q_2$ (marked as +) for $\Omega=5$ and $\Theta=7/16$. Simulated data presented are using $q_1=0.8$ ($q_2=0.1$) when varying $q_2(q_1)$ - qualitatively similar figures can be obtained for other values of $q_{1,2}$.} 
  \label{fig_frac_ratio}
\end{figure}

For an intermediate overall density, the plus-type and/or minus-type
particles form shocks near the junctions between sections. In
contrast to low overall densities where $F^b_+$ is highly dependent
on $q_1$ and weakly dependent on $q_2$, in the intermediate overall
densities, the fraction in the bipolar section $F^b_+$ shows
relatively weak dependence on both switching rates $q_{1,2}$. The
left panel in Figure~\ref{fig_frac_ratio} shows as an example with
an overall density $\Theta=7/16$ of how the fraction in occupancy
varies changes with the rates $q_{1,2}$.

\subsection{Dominance in current}\label{sec_unid_curr}
\begin{figure}
  \centering
  \includegraphics[width=10cm]{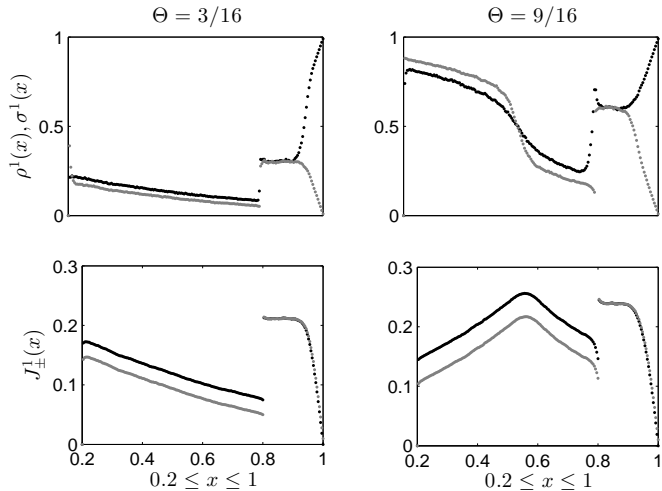}
  \caption{Numerical current profiles (bottom) with corresponding density profiles (top) for overall density $\Theta=3/16$ (left) and $\Theta=9/16$ (right). Other parameters use $\Omega=5,~q_1=0.8,~q_2=0.4$. Dark dots are for plus-type particles while gray dots are for the minus-type.}
  \label{fig_den_curr}
\end{figure}

The directed currents $J_{\pm}$ are also important quantities to
characterize the transport capacity - a large current indicates efficient transport. In contrast to the standard
unidirectional ASEP on a single lane (where both the density and
current are constant along the bulk of the lane), in our model,
neither the density nor the directed current of each type of
particles on a single track is constant; see Figure~\ref{fig_den_curr} for examples. Thus, we consider the
average unidirectional current
$$
\langle J_{\pm}\rangle_x:=\frac{\int^1_{0}J_{\pm}(x)dx}{1-x_1}
$$
and define an overall
current $\langle J\rangle_x:=\langle J_+\rangle_x+\langle
J_-\rangle_x$. The dominance in current of particles can be investigated by looking at the ratio of average current for plus-type particles:
$$
R_{+}:=\frac{\langle J_+\rangle_x}{\langle J\rangle_x}=\frac{\langle J_+\rangle_x}{\langle J_+\rangle_x+\langle J_-\rangle_x}.
$$
We simultaneously define the ratio of average currents within each section. In the unipolar sections, plus- and minus-type particles equally contribute to the average currents as the net current is zero. Thus the ratio in the entire bundle $R_+$ depends only on the ratio within the bipolar section,
$$R^b_+:=\frac{\int_{x_1}^{x_2}J_+(x)dx}{\int_{x_1}^{x_2}(J_+(x)+J_-(x))dx}.$$

When the density profiles in the bipolar section are in the LL phase
($\rho,\sigma\ll 1/2$), the dominance in current is equivalent to
that in occupancy, $R^b_+\approx F^b_+$ (seen from Figure~\ref{fig_den_curr}~(left panel)). However, if shocks form in
the bipolar section, higher densities (above one half) can give a
lower current (seen from Figure~\ref{fig_den_curr}~(right panel)). Thus dominance in occupancy does not imply dominance
in current; see Figure~\ref{fig_frac_ratio} as an example. Moreover,
by comparing the two panels in Figure~\ref{fig_frac_ratio}, $R^b_+$
shows a larger range than $F^b_+$ on changing parameters. Thus, for
an intermediate overall density, one could say that current plays a
more important role than occupancy in determining transport of particles. In addition, we can see that the
ratio in current $R^b_+$ shows a larger range on changing $q_1$ than
on changing $q_2$ in this example. This suggests that $q_1$ is more
important than $q_2$ for the dominance.

\section{Discussion}\label{sec_dis}
In this paper, we introduce an ASEP-type model to describe the bidirectional motility of particles on an oriented bundle of tracks. This aims to model the motion of cargos/motor complexes undergoing transport along an antipolar MT bundle within a cell. Our model is certainly a great simplification of cell transport processes. It is parameterized by the particle turning rate $\Omega$ (the inverse of the run length), the overall density $\Theta$ (the proportion of sites that are occupied by particles) and the obstacle/end-induced switching rates $q_{1,2}$ (i.e. the rate at which particles switch MTs at the junctions between the unipolar and bipolar sections). We use numerical simulations and mean-field approximation to investigate the dependence of the stationary density profiles within the bundle on these parameters.

We observe that, as expected, the switching rates $q_{1,2}$ have a major effect on the distribution of particles along the bundle. Although we have not fully explored the dependence of phases on parameters we highlight below a number of interesting features about the system. Even for {\em low} overall density $\Theta$, particles (cargo/motor complexes) can queue to form accumulations at minus ends in addition to any plus-end queuing observed in simpler situations \cite{Ashwin2010}. The critical value of the end-induced switching rate $q_2$ above which shocks are formed at minus ends is investigated in Section~\ref{sec_low}. For {\em intermediate} overall density $\Theta$, we find a variety of phases, including a new ``smooth connection'' phase where in the limit, the density profile of one type of particle on a track smoothly passes through one half while the other type displays a shock. We discuss a variety of other phases and have investigated the role of the switching rates $q_{1,2}$ in determining which motor is dominant in the transport. For {\em low} overall densities, the obstacle-induced switching rate $q_1$ influence the fraction of particles of one type within the bipolar section in an approximately linear way. For {\em high} overall density $\Theta$, our study shows that although the fraction of occupancy by different type particles does not vary much with rates $q_{1,2}$, the fraction that actually contributes to the current (i.e., transport) from different types may vary much more.

Our model is inspired by {\em in vivo} experimental observations, although the model has been simplified in many ways as we now discuss. For example, it is probably that hopping between MTs is not restricted to the ends of the bipolar section, but may occur throughout the bipolar section \cite{Schuster2011c}. This would allow more possibilities for transition events, and the possibility of plus-type particles (in addition to minus-type particles) switching MTs. A switch in transport direction can result from hopping between MTs and the activity of kinesin-3 alone or it could be a consequence of dynein binding to the cargo, which in case of EEs was shown to override kinesin-3 activity \cite{Schuster2011b}. Our model also assumes only two lanes in each motility track. Moreover, it is known that numerous MTs form a bundle \cite{Schuster2011c}, each consisting of 13 protofilaments \cite{Tilney1973}. Thus, many more tracks might support bidirectional motility of the cargo/motor complexes and allowing opposite-directed particles moving on the same protofilament will certainly give new effects such as increasing the jamming at MT plus ends, as discussed in \cite{Lin2011}. The complex geometry of the bundle will also contribute to new effects of cooperative transport, already considered for unidirectional transport in \cite{Neri2011}.

Notwithstanding these simplifications, we suggest the model could be useful in a number of ways, especially when improved measurement of transition rates is possible in vivo, and we highlight some of these below:
\begin{enumerate}
\item For low overall density of particles, it is possible to have an accumulation of particles at minus ends of MTs. As {\em in vivo} experiments so far show no obvious accumulation of early endosome cargos throughout the entire cell \cite{Schuster2011c}, if this model is accurate then it suggests the end-induced switching rate $q_2$ must be high {\em in vivo}. There are clearly other possible explanations due to features not included in the model such as a high turning rate of minus-directed organelles at minus ends (which might be due to an accumulation of motors at the minus end taking dynein to plus ends) could avoid such an accumulation of EEs; this would be a similar mechanism to that suggested in \cite{Ashwin2010,Schuster2011a} where an accumulation of dynein motors increases the turning rate of EEs near plus-ends and so avoids accumulations of EEs.
\item For low overall density of cargos, the fraction of occupancy can be used to show how the proportion of EEs carried by kinesin-3 among all EEs (carried by either kinesin-3 or dynein) varies with the parameters in Section~\ref{sec_domin}. These estimates could in principle be explored experimentally to test the modelling assumptions used to describe {\em in vivo} bidirectional transport.
\end{enumerate}
Further experimental work will improve our qualitative and quantitative knowledge of the details of transport process in the living cell, though the consequences of individual cargo behaviour on coordinated transport are not always clear. By using the model discussed in this paper, and developments thereof, there is a platform with which one can address various open questions about the nature and function of the coordinated systems involved in cell transport

\appendix
\section{General density profiles in the bipolar section}\label{App_den}
In the bipolar section, the net current $J$ (as discussed in Section~\ref{sec_meanfield}) given by $J=\rho(1-\rho)-\sigma(1-\sigma)$ is constant when the system is in a statistically stationary state. This density-current relation gives
$$
\sigma=\frac{1}{2}\pm\frac{2\rho-1}{2}\left(\sqrt{\frac{4J}{(2\rho-1)^2}+1}-1\right)\pm\frac{2\rho-1}{2},
$$
which implies the ``equal'' and ``complementary'' density solutions: 
$$
\sigma=\rho+\frac{2\rho-1}{2}\left(\sqrt{\frac{4J}{(2\rho-1)^2}+1}-1\right), ~~\sigma=1-\rho-\frac{2\rho-1}{2}\left(\sqrt{\frac{4J}{(2\rho-1)^2}+1}-1\right).
$$
Thus, the first order ODE (\ref{eq_pdesimple}) from the mean field
approximation reads
\begin{equation}\label{eq_genODE}
0 = \frac{d\rho}{dx} +
\frac{\Omega}{2}\left(\pm\sqrt{\frac{4J}{(2\rho-1)^2}+1}-
1\right)\nonumber
\end{equation}
with a general solution of ``equal'' or ``complementary'' density
\begin{equation}\label{eq_sol}
\left(\left(1+\frac{4J}{(2\rho-1)^2}\right)^{3/2}\pm
1\right)(2\rho-1)^3=\mp12J\Omega x+C
\end{equation}
for any non-zero net current $J$ on taking options in ``$\pm$'' and ``$\mp$''.

\section{Mean-field analysis for an LL-SL phase}\label{App_MFlow}
When both types of particles in
the bipolar section are in low densities, they are dominated by the
injection rates $\alpha^b_{\pm}$ which are approximated by
\begin{equation}\label{eq_BC_alpha_sigma}
\alpha^b_{+}=\bar{\sigma}q_1\mbox{~and~}\alpha^b_{-}=\bar{\sigma}(1-q_1).
\end{equation}
where $\bar{\sigma}$ is the equal density in the unipolar section.
Hence from (\ref{eq_rho_smallJ}) and (\ref{eq_sigma_smallJ}), the densities on the first track in the bipolar section are approximated by
\begin{equation}\label{eq_den_LL}
\rho=-J\Omega (x-x_1)+\alpha^b_{+},~~\sigma=-J\Omega (x-x_2)+\alpha^b_{-}.
\end{equation}
Together with the approximation $J=\rho-\sigma$, we find
\begin{equation}\label{eq_J_LL}
J=\frac{\alpha^b_{+}-\alpha^b_{-}}{\Omega(x_2-x_1)+1}=\frac{\bar{\sigma}(2q_1-1)}{\Omega(x_2-x_1)+1},
\end{equation}
which gives a zero net current if $q_1=1/2$. The density
expressions~(\ref{eq_den_LL}) together with the symmetry
$\rho^1(x)=\rho^2(1-x)$ and $\sigma^1(x)=\sigma^2(1-x)$ give the following
$$\rho^1(x)+\sigma^2(x)=\rho^2(x)+\sigma^1(x)=\bar{\sigma},~x\in(x_1,x_2).$$
Thus
\begin{equation}\label{eq_theta}
\Theta=\frac{\int_{x_1}^1\rho(x)+\sigma(x)dx}{2(1-x_1)}=\frac{\bar{\sigma}/\Omega+(1-2\bar{\sigma}/\Omega)\bar{\sigma}}{2(1-x_1)}.
\end{equation}
That is, given a low overall density $\Theta$, a sufficiently large $q_2$ and assuming densities are low in the bipolar section, we have
\begin{equation}\label{eq_thetatobar}
\bar{\sigma}=\frac{1+\Omega}{4}-\sqrt{\frac{(1+\Omega)^2}{16}-(1-x_1)\Omega\Theta}.
\end{equation}
Substituting (\ref{eq_thetatobar}), (\ref{eq_J_LL}) and (\ref{eq_BC_alpha_sigma})
into (\ref{eq_den_LL}) gives the analytical approximation of the
density profiles in the bipolar section in an LL-SL phase. This,
together with density profiles in the unipolar sections, gives the
density profiles in the entire bundle.

\bibliographystyle{plain}

\end{document}